\documentclass[prl,twocolumn,showpacs]{revtex4}
\usepackage{graphicx}
\begin{document}

\title{Second Generation of Moore--Read Quasiholes 
       in a Composite Fermion Liquid}

\author{
   Arkadiusz W\'ojs,$^{1,2}$
   Daniel Wodzi\'nski,$^1$ and 
   John J. Quinn$^2$}

\affiliation{
   \mbox{
   $^1$Institute of Physics, Wroclaw University of Technology,
       Wybrze\.ze Wyspia\'nskiego 27, 50-370 Wroclaw, Poland}\\
   $^2$Department of Physics, University of Tennessee, 
       Knoxville, TN 37996, USA}

\begin{abstract}
Two- and three-body correlations of incompressible quantum 
liquids are studied numerically.
Pairing of composite fermions (CFs) in the 1/3-filled second 
CF Landau level is found at $\nu_e=4/11$. 
It is explained by reduced short-range repulsion due to ring-like 
single-particle charge distribution.
Although Moore--Read state of CFs is unstable in the 1/2-filled 
second CF level, condensation of its quasiholes is a possible 
origin of incompressibility at $\nu_e=4/11$.
Electron pairing occurs at $\nu_e=7/3$ and $13/3$, but with 
different pair--pair correlations.
Signatures of triplets are found at higher fillings.
\end{abstract}
  
\maketitle

Strong magnetic field $B$ applied to a two-dimensional 
electron gas (2DEG) rearranges its single-particle density 
of states to a series of discrete Landau levels (LL$_n$).
When the cyclotron gap $\hbar\omega_c\propto B$ exceeds 
Coulomb energy $e^2/\lambda\propto\sqrt{B}$ ($\lambda=
\sqrt{\hbar c/eB}$ being the magnetic length), the low-energy 
dynamics depends on interactions in one, partially filled LL.
Despite reminiscence to atomic physics, macroscopic 
degeneracy and a distinct scattering matrix lead to very 
different, fascinating behavior \cite{Yoshioka02}.

Fractional quantum Hall effect \cite{Tsui82} reveals plethora 
of highly correlated electron phases at various LL filling 
factors $\nu_e=2\pi\varrho\lambda^2$ ($\varrho$ being the 2D 
concentration).
Among them are Laughlin \cite{Laughlin83} and Jain \cite{Jain89} 
incompressible liquids (IQLs) with fractionally charged 
quasiparticles (QPs) at $\nu_e={1\over3}$ or ${2\over5}$, 
Wigner crystals \cite{Lozovik75} at $\nu_e\ll1$, and stripes 
\cite{Koulakov96} in high LLs.
Besides transport \cite{Tsui82}, they are probed by shot-noise 
(allowing detection of fractional charge of the QPs 
\cite{Saminadayar97}) and optics (with discontinuities in 
photoluminescence energy related to the QP interactions 
\cite{Goldberg90}).

A key concept in understanding IQLs is Jain's composite 
fermion (CF) picture \cite{Jain89}.
The CFs are fictitious particles, electrons that captured 
part of the external magnetic field $B$ in form of 
infinitesimal tubes carrying an even number $2p$ of flux 
quanta $\phi_0=hc/e$.
The most prominent IQLs at $\nu_e=n(2ps\pm1)^{-1}$ are 
represented by the completely filled $s$ lowest LLs of the 
CFs (CF-LL$_n$ with $n<s$) in a residual magnetic field 
$B^*=B-2p\phi_0\varrho$.

Not all IQLs are so easily explained by the CF model, 
e.g., Haldane--Rezayi \cite{Haldane88} and Moore--Read 
\cite{Moore91} paired liquids proposed for $\nu_e={5\over2}$.
Because of nonabelian statistics of its quasiholes (QHs), 
especially the latter state has recently stirred renewed 
interest as a candidate for quantum computation in a solid-state 
environment \cite{DasSarma05}.

Another family of IQLs discovered by Pan {\em et al.} \cite{Pan03} 
at $\nu_e={4\over11}$, ${3\over8}$, and ${5\over13}$ corresponds 
to fractional CF fillings $\nu_{\rm CF}=\nu_e(1-2p\nu_e)^{-1}=
{4\over3}$, ${3\over2}$, and ${5\over3}$ (with $p=1$).
Assuming spin polarization, all these states have a partially 
filled CF-LL$_1$.
Their incompressibility results from residual CF--CF interactions. 
Familiar values of $\nu_{\rm CF}$ suggested similarity between 
partially filled electron and CF LLs \cite{Mandal02}.
For $\nu_e={4\over11}$ and ${5\over13}$, it revived the ``QP 
hierarchy'' \cite{Haldane83}, whose CF formulation consists 
of the CF$\,\rightarrow\,$electron mapping followed by 
reapplication of the CF picture in CF-LL$_1$, leading to 
a ``second generation'' of CFs \cite{Smet03}.
However, this idea ignored the requirement of a strong 
short-range repulsion \cite{Haldane85,hierarchy}.
Indeed, it was later excluded in exact diagonalization 
studies \cite{clusters}, in which a different series of 
finite-size $\nu_e={4\over11}$ liquids with larger gaps was 
identified.
On the other hand, Moore--Read liquid of paired CFs was 
tested \cite{Park98} for $\nu_e={3\over8}$, but it was 
eventually ruled out in favor of the 
stripe order \cite{Lee01,Shibata04}.

In this Letter, we study two- and three-body correlations 
in several IQLs whose origin of incompressibility remains 
controversial.
We find evidence for CF pairing in the $\nu_e={4\over11}$ 
liquid, hence interpreted as {\em a condensate of (nonabelian) 
QHs of the ``second generation'' Moore--Read state of the CFs}.
The pair--pair or QH--QH correlations are not defined, but 
a Laughlin form is excluded. 

\begin{figure}
\includegraphics[width=3.4in]{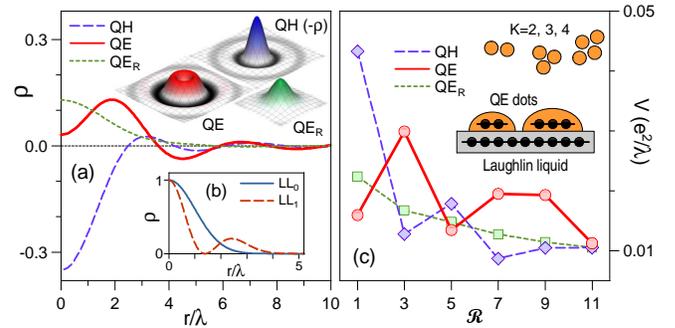}
\caption{(color online).
   (a) Radial charge distribution profiles of 
   different composite fermions: Laughlin quasielectron 
   (QE), quasihole (QH), and reversed-spin quasielectron 
   (QE$_{\rm R}$); results obtained from exact 10-electron
   diagonalization; $\lambda$ is the magnetic length.
   (b) Same for electrons in two lowest Landau levels.
   (c) Haldane pseudopotentials (interaction energy $V$ 
   vs.\ relative pair angular momentum $\mathcal{R}$) 
   for composite fermions; inset: schematic of ``artificial 
   composite fermion atoms.''}
\label{fig1}
\end{figure}

In Fig.~\ref{fig1}(a,b) charge-density distributions of 
electrons are compared with three different CF quasiparticles 
at $\nu_e={1\over3}$.
Laughlin liquid is a filled spin-polarized CF-LL$_0$, and 
its quasielectron (QE), quasihole (QH), and reversed-spin 
quasielectron (QE$_{\rm R}$) correspond to a particle in 
CF-LL$_1$, a vacancy in CF-LL$_0$, and a spin-flip particle 
in CF-LL$_0$, respectively.
Particles/holes in CF-LL$_0$ resemble those in LL$_0$.
However, the ring structure in CF-LL$_1$ makes the QEs 
different from the electrons and causes strong reduction of 
the QE--QE repulsion at short range [cf.\ Fig.~\ref{fig1}(c)].
Such interaction cannot \cite{Haldane85,hierarchy}
produce a Laughlin IQL of the QEs at the $\nu={1\over3}$ 
filling of CF-LL$_1$.
Instead, other QE--QE correlations must be considered.

Spontaneous QE cluster formation would be somewhat analogous 
to the self-assembled growth of strained quantum dots 
\cite{qdots}.
A full CF-LL$_0$ representing the uniform-density Laughlin 
liquid plays the role of a ``wetting layer.'' 
Over this background, in analogy to atoms grouping into dots 
to minimize the elastic energy, QEs moving within CF-LL$_1$ 
arrange themselves into pairs or larger QE clusters easily 
pinned down by disorder.
While in electronic ``artificial atoms'' the self-organization 
of real atoms serves a purpose of external confinement for 
the electrons, in their CF analogs both these roles are 
played by the QEs.
Another distinction is the fractional charge of bound QE 
carriers.
A similar {\em electron}--atom analogy was earlier explored
for condensation of cold atoms \cite{Wilkin00}. 

In numerics we considered $N\le12$ particles 
($N=12$ being divisible by $K=2$, 3, and 4) of charge $q$ 
($-e$ for electrons and $-{1\over3}e$ for the CFs) confined 
to a Haldane sphere \cite{Haldane83} of radius $R$.
For its high symmetry, this geometry is especially useful in 
studying quantum liquids.
The radial magnetic field $B$ is created by 
a Dirac monopole of strength $2Q=4\pi R^2 B\phi_0^{-1}$.
The single-particle LLs are distinguished by shell angular 
momentum $l\ge Q$.

As for a partially filled atomic shell, the many-body 
hamiltonian on a sphere is determined by particle number $N$, 
shell degeneracy $g=2l+1$, and interaction matrix elements.
Using Clebsch-Gordan coefficients, the latter are related to
Haldane \cite{Haldane87} pseudopotentials $V_L$ (energies of 
pairs with angular momentum $L$).
The pseudopotential combines information about the potential 
$v(r)$ and shell wavefunctions, so it may not be similar in 
different systems with the same (Coulomb) forces.
In macroscopic quantum Hall systems, only the ratio $\nu=N/g$
(filling factor) is important, and $V$ is a function of relative 
pair angular momentum $\mathcal{R}=2l-L$ (for fermions, an odd 
integer).
The strategy in exact diagonalization is therefore to study 
different finite systems $(N,2l)$ with a realistic interaction 
$V(\mathcal{R})$, in search of those properties which scale 
properly with size and persist in the macroscopic limit.

In the following we will distinguish $\nu_e=2\pi\varrho\lambda^2$ 
from the effective filling factor $\nu=N/g<1$ of only those 
electrons or CFs in their highest, partially filled shell. 
In LL$_n$, $\nu_e=2n+\nu$.
In CF-LL$_n$ (assuming spin-polarization) $\nu_{\rm CF}=n+\nu$ 
and $\nu_e=\nu_{\rm CF}(2p\nu_{\rm CF}+1)^{-1}$.

The CF pseudopotentials shown in Fig.~\ref{fig1}(c) were 
obtained using a similar method to Ref.~\cite{Lee01}, by 
combining short-range data from exact diagonalization 
\cite{hierarchy} with long-range behavior of point charges 
$\pm{1\over3}e$.
Weak QE--QE repulsion at $\mathcal{R}=1$ is the reason why the 
$\nu={1\over3}$, ${2\over3}$, and ${1\over2}$ states of QEs 
are not the ``second generation'' Laughlin, Jain, or Moore--Read 
states (of QEs).
The average QE--QE interaction energies (per particle) in these 
states overestimates the actual QE eigenenergies by at least 
$0.003\,e^2/\lambda$ (6--7\%).
Clearly, the microscopic origin of the observed QE 
incompressibility must be different.

What are these known correlations, excluded for QEs? 
Laughlin correlations result from strong short-range repulsion 
(such as between electrons in LL$_0$).
They consist of the maximum avoidance of pair states with the 
smallest $\mathcal{R}$.
E.g., Laughlin $\nu={1\over3}$ state is the zero-energy ground 
state of a model pseudopotential $V=\delta_{\mathcal{R},1}$
\cite{Haldane83}.
For more realistic interactions, the exact criterion is that
$V$ must rise faster than linearly when $\mathcal{R}$ decreases 
\cite{hierarchy}.
A linear decrease of $V$ between $\mathcal{R}=1$ and 5 (such 
as in LL$_1$) leads to different correlations.
E.g., Moore--Read $\nu={1\over2}$ liquid involves pairing and 
Laughlin correlations among pairs.
It is the zero-energy ground state of a model three-body 
pseudopotential $V=\delta_{\mathcal{T},3}$ ($\mathcal{T}=3l-L\ge3$ 
is the relative triplet angular momentum, proportional to the 
area spanned by three particles) \cite{Moore91}.

Weak QE--QE repulsion at $\mathcal{R}=1$ compared to $\mathcal{R}
=3$ could force QEs into even larger clusters.
As a simple classical analogy, consider a string of point particles, 
one per unit length, with a repulsive potential $v_a(r)=a+(1-a)r$ 
for $r<1$ and $1/r^2$ otherwise.
Equal spacing is favored for $a>1.64$, and transitions to pairs, 
triplets, and larger clusters occur for decreasing $a$.
A similar rearrangement might occur when going from LL$_0$ to 
LL$_1$ and CF-LL$_1$, with $V(1)$ playing the role of $v_a(0)
\equiv a$.

\begin{figure}
\includegraphics[width=3.4in]{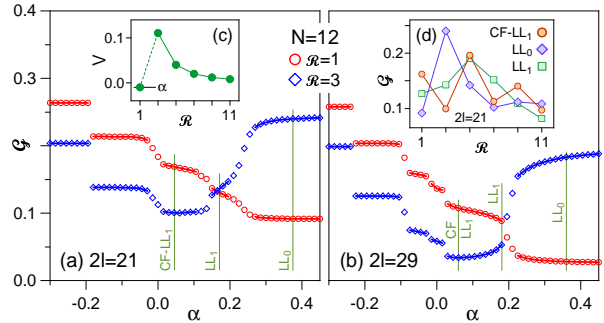}
\caption{(color online).
   Haldane pair amplitudes $\mathcal{G}$ ($\sim\,$number of pairs) 
   at relative pair angular momenta $\mathcal{R}=1$ and 3, 
   of $N=12$ fermions in angular momentum shells with $2l=21$ (a) 
   and $2l=29$ (b), as a function of parameter $\alpha$ of the 
   interaction pseudopotential shown in (c).
   (d) Amplitudes $\mathcal{G}(\mathcal{R})$
   of electrons and composite fermions in different Landau 
   levels.}
\label{fig2}
\end{figure}

In Fig.~\ref{fig2} we plot two leading ``Haldane amplitudes'' 
\cite{Haldane87}, $\mathcal{G}(1)$ and $\mathcal{G}(3)$.
The discrete pair-correlation function $\mathcal{G}(\mathcal{R})$ 
is proportional to the number of pairs with a given $\mathcal{R}$ 
and normalized to $\sum_\mathcal{R}\mathcal{G}(\mathcal{R})=1$.
It connects many-body interaction energy with a pseudopotential, 
$E={N\choose2}\sum_\mathcal{R}\mathcal{G}(\mathcal{R})V(\mathcal{R})$.
Here, $\mathcal{G}$ is calculated in the ground states of $N=12$ 
particles at $2l=21$ and 29 (corresponding to $\nu={1\over2}$ 
and ${1\over3}$ for the QEs \cite{clusters}) with model 
interaction shown in the inset: $V_\alpha(1)=\alpha$ and 
$V_\alpha(\mathcal{R}>1)=1/\mathcal{R}^2$.
At $\alpha>0.3$, $\mathcal{G}(1)$ takes on the minimum possible 
value, which means Laughlin correlations (no clusters).
At $\alpha<-0.25$, $\mathcal{G}(1)$ reaches maximum, and the 
particles form one big $\nu=1$ quantum Hall droplet (QHD).
The transition between the two limits occurs quasi-discontinuously
through a series of well-defined states seen as plateaus in 
$\mathcal{G}(\alpha)$.

The cluster size $K$ cannot be assigned to each state because
the number of plateaus depends on the choice of $V_\alpha$.
The comparison of $\mathcal{G}(1)$ with the values predicted 
for $N/K$ independent QHDs of size $K=2$, 3, and 4 is not
convincing because in a few-cluster system each QHD is relaxed 
by the cluster--cluster interaction, lowering $\mathcal{G}(1)$.
Another problem is the contribution to $\mathcal{G}(1)$ from 
pairs of particles belonging to different clusters.
Nevertheless, it is clear that the ``degree of clustering'' 
changes as a function of $\alpha$ in a quantized fashion, 
supporting the picture of $N$ particles grouping into various 
clustered configurations.
Furthermore, the values of $\alpha$ for which $V_\alpha$ reproduces 
the exact ground states of QEs or electrons belong to different 
continuity regions, confirming different correlations in 
LL$_0$, LL$_1$, and CF-LL$_1$ (except for a possible similarity
of the $\nu={1\over3}$ states in LL$_1$ and CF-LL$_1$).

\begin{figure}
\includegraphics[width=3.4in]{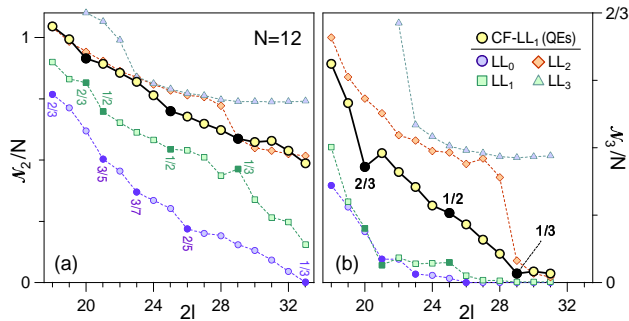}
\caption{(color online).
   Number of pairs $\mathcal{N}_2$ (a) and triplets 
   $\mathcal{N}_3$ (b) with the minimum relative angular
   momentum ($\mathcal{R}=1$ or $\mathcal{T}=3$) for $N=12$ 
   electrons or composite fermions in angular momentum shells 
   with $2l=18$ to 33, corresponding to fractional Landau level
   fillings ${1\over3}\le\nu\sim N/(2l+1)\le{2\over3}$.
   Finite-size incompressible states are labeled by $\nu$.}
\label{fig3}
\end{figure}

In Fig.~\ref{fig3}(a) we compare $\mathcal{N}_2={N\choose2}\,
\mathcal{G}(1)$, the number of pairs with $\mathcal{R}=1$,
calculated in the ground states of $N=12$ CFs and electrons 
as a function $2l$.
The downward cusps in $\mathcal{N}_2(2l)$ at a series of 
Laughlin/Jain states in LL$_0$ are well understood.
We also marked $2l=2N-3$ and $3N-7$ corresponding to 
incompressible $\nu={1\over2}$ and ${1\over3}$ ground 
states in LL$_1$ and CF-LL$_1$ \cite{clusters}, and their 
particle-hole conjugates ($N\rightarrow g-N$) at $2l=2N+1$ 
and ${3\over2}N+2$.

The comparison of $\mathcal{N}_2$ tells about short-range pair 
correlations in different LLs.
There are significantly more pairs in CF-LL$_1$ and in excited 
electron LLs than in LL$_0$.
In LL$_1$, the Moore--Read state is known to be paired, and 
indeed $\mathcal{N}_2\approx{1\over2}N$ at $\nu={1\over2}$.
A similar value is obtained for the (not well understood) 
$\nu={1\over3}$ state at $2l=29$.
The CF-LL$_1$ is different (in terms of $\mathcal{N}_2$) from 
LL$_0$ or LL$_1$ in the whole range of $18\le2l\le33$.
However, it appears similar to LL$_2$ at both $2l\le23$ and 
$2l\ge29$.
Also, LL$_2$ and LL$_3$ look alike for $23\le2l<29$.
While convincing assignment of $\nu$ to a finite state $(N,2l)$ 
requires studying size dependence (we looked at different $N\le12$), 
notice that $N/g={1\over2}$ at $2l=23$, and $2l=29$ is the $\nu=
{1\over3}$ state in LL$_1$ and CF-LL$_1$.
Note also that similar short-range correlations do not guarantee 
high wavefunction overlaps.
Here, only $\left<{\rm LL}_2|{\rm LL}_3\right>^2$ reaches 0.67 
while all other overlaps, including $\left<{\rm QE}|{\rm LL}_n
\right>^2$, essentially vanish.

In Fig.~\ref{fig3}(b) we plot $\mathcal{N}_3$, the number of
``compact'' triplets with $\mathcal{T}=3$.
It is proportional to the first triplet Haldane amplitude 
and tells about short-range three-body correlations.
In both LL$_0$ and LL$_1$, $\mathcal{N}_3$ decreases roughly
linearly as a function of $2l$ and drops to essentially zero 
at $2l=21$, the smallest value at which the $\mathcal{T}=3$ 
triplets can be completely avoided.
Exactly $\mathcal{N}_3=0$ would indicate the Moore--Read state, 
but its accuracy for the actual $\nu={1\over2}$ ground state 
in LL$_1$ depends sensitively on the quasi-2D layer width and 
on the surface curvature.
Nevertheless, clusters larger than pairs clearly do not form 
in neither LL$_0$ nor LL$_1$ at $\nu\le{1\over2}$.

The number of QE triplets in CF-LL$_1$ is also a nearly 
linear function of $2l$, but it drops to zero at $2l=3N-7=29$,
earlier identified with $\nu={1\over3}$ in this shell 
(i.e. with $\nu_e={4\over11}$) \cite{clusters}.
In connection with having $\mathcal{N}_2\approx{1\over2}N$ pairs, 
{\em vanishing of $\mathcal{N}_3$ is the evidence for QE 
pairing at $\nu_e={4\over11}$}.

The elementary excitations that appear in the paired $\nu={1\over2}$ 
Moore--Read state when $2l>2N-3$ are the ${1\over4}q$-charged QHs 
(of the Laughlin liquid of pairs) and pair-breaking neutral-fermion 
excitations \cite{Moore91}.
Being paired, the QE state at $2l=3N-7$ {\em can only contain the 
QHs but no pair-breakers}.
The interaction of Moore--Read QHs in CF-LL$_1$ is not known, 
but evidently it causes their condensation into an incompressible
liquid at $\nu={1\over3}$.

The ``second generation'' (to distinguish from $\nu_e={5\over2}$) 
Moore--Read state of QEs would occur at $\nu={1\over2}$ in CF-LL$_1$ 
(i.e., at $\nu_{\rm CF}={3\over2}$ or $\nu_e={3\over8}$).
Its instability \cite{Lee01,Shibata04} does not preclude reentrance 
with additional QHs at a lower $\nu$ and, in particular, their 
condensation at $\nu={1\over3}$ (i.e., at $\nu_{\rm CF}={4\over3}$ 
or $\nu_e={4\over11}$).
A similar situation occurs with Jain $\nu={2\over7}$ state, obtained 
(in Haldane hierarchy) from Laughlin $\nu={1\over3}$ state by addition 
of ``second generation'' Laughlin QHs.
There, stability of the $\nu={2\over7}$ daughter does {\em not} require 
stability of the $\nu={1\over3}$ parent.

The value of $2l=3N-7$ precludes a Laughlin state of pairs (or, 
equivalently, of the QHs).
To show it, let us use the following pictorial argument, equivalent 
to a more rigorous derivation.
Laughlin $\nu={1\over3}$ state (of individual particles) can 
be viewed as $\bullet$$\circ$$\circ$$\bullet$$\dots$$\bullet$%
$\circ$$\circ$$\bullet\equiv(\bullet$$\circ$$\circ$)$\bullet$, 
with ``$\bullet$'' and ``$\circ$'' denoting particles and vacancies.
Counting the total LL degeneracy $g$ leads to the correct value 
of $2l=3N-3$.
The Moore--Read state, i.e., the Laughlin state of pairs at 
$\nu={1\over2}$, is represented by $(\bullet$$\bullet$$\circ$%
$\circ)$$\bullet$$\bullet$, yielding $2l=2N-3$.
A Laughlin state of pairs at $\nu={1\over3}$ would correspond 
to $(\bullet$$\bullet$$\circ$$\circ$$\circ$$\circ$)$\bullet$%
$\bullet$, predicting (incorrectly) $2l=3N-5$.
Assumming pairing, $2l=3N-7$ can only be obtained using a 
two-pair unit cell, $(\bullet$$\bullet$$\circ$$\circ$$\bullet$%
$\bullet$$\circ$$\circ$$\circ$$\circ$$\circ$$\circ)$$\bullet$%
$\bullet$$\circ$$\circ$$\bullet$$\bullet$, corresponding to 
more complicated pair--pair correlations.

At higher fillings of CF-LL$_1$, $\mathcal{N}_3\approx{1\over3}N$ 
at $2l=20$ suggests division of $N$ QEs into ${1\over3}N$ 
triplets at $\nu={2\over3}$, and $\mathcal{N}_3\approx
{1\over6}N$ at $2l=25$ implies a more complicated cluster 
configuration (with mixed sizes) at $\nu={1\over2}$.
LL$_2$ and LL$_3$ look alike (and different from LL$_0$ or 
CF-LL$_1$) at $23\le2l<29$, both having $\mathcal{N}_3\approx
{1\over3}N$.
At $2l=29$, $\mathcal{N}_3$ for LL$_2$ drops rapidly to almost
zero.
This further supports similarity of the $\nu={1\over3}$ states 
in LL$_2$ and CF-LL$_1$.

\begin{figure}
\includegraphics[width=3.4in]{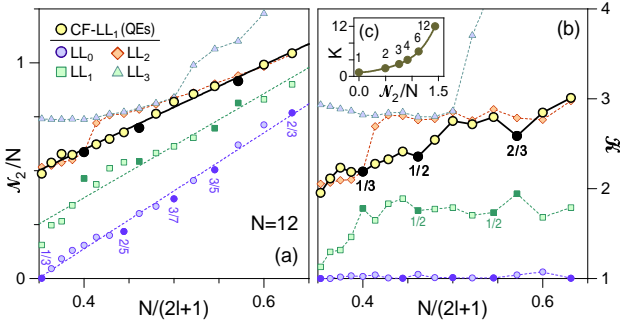}
\caption{(color online).
   Number of pairs $\mathcal{N}_2$ with the minimum relative 
   angular momentum $\mathcal{R}=1$ (a) 
   and estimated average cluster size $\mathcal{K}$ (b)
   for $N=12$ electrons or composite fermions in Landau 
   levels angular momentum shells with $2l=18$ to 33, 
   plotted as a function of the filling factor 
   $\nu\sim N/(2l+1)$.}
\label{fig4}
\end{figure}

In Fig.~\ref{fig4}(a) we replot $\mathcal{N}_2$ as a 
function of $N/g\sim\nu$.
The quasi-linear dependences for LL$_0$, LL$_1$, and 
CF-LL$_1$ all aim correctly at $\mathcal{N}_2=2N-3$ for
$\nu=1$, but start from different values, $\mathcal{N}_2
\approx0$, ${1\over4}N$, and ${1\over2}N$, at $\nu={1\over3}$.
Regular dependence allows subtraction from $\mathcal{N}_2$ 
the contribution from pairs belonging to different clusters.
As reference we used ground states of $V=\delta_{\mathcal{R},1}$.
This short-range repulsion guarantees maximum avoidance 
of $\mathcal{R}=1$; its $\mathcal{N}_2^*$ contains only 
the inter-cluster contribution.
To compare $\mathcal{N}_2$ of QEs or electrons with 
$\mathcal{N}_2^*$, we:
(i) calculated $\mathcal{N}_2$ for a single $K$-size cluster, 
and multiplied it by $N/K$ to obtain relation between 
$\mathcal{N}_2$ and $K$ in an idealized clustered state of 
$N$ particles, 
(ii) using this relation [cf.\ Fig.~\ref{fig4}(c)], 
converted $\mathcal{N}_2$ and $\mathcal{N}_2^*$ into 
the (average) cluster sizes $K$ and $K^*$;
(iii) defined $\mathcal{K}=K-(K^*-1)$ as the cluster size 
estimate free of the inter-cluster contribution.

The result in Fig.~\ref{fig4}(b) indicates pairing in LL$_1$ 
at ${1\over3}\le\nu\le{2\over3}$, and in both CF-LL$_1$ and 
LL$_2$ at $\nu\le{1\over3}$.
Triplets seem to form in CF-LL$_1$ at $\nu={2\over3}$, in 
LL$_2$ at ${1\over3}\le\nu\le{2\over3}$, and in LL$_3$ at 
$\nu\le{1\over2}$.
The $\nu={1\over2}$ state of QEs falls between $\mathcal{K}
=2$ and 3, suggesting mixed-size clusters.

In conclusion, we studied two- and three-body correlations 
of several quantum liquids.
We found evidence for pairing of CFs at $\nu_e={4\over11}$ and
interpret this state as a condensate of ``second generation'' 
Moore--Read QHs.

AW thanks Wei Pan and Jan Jaroszy\'nski for helpful discussions.
Work supported by DOE Basic Energy Science and Grant 
2P03B02424 of the Polish MENiS.

\end{document}